\documentstyle[epsf,rotate,twocolumn,aps]{revtex}

\begin{document}
\draft

\twocolumn[\hsize\textwidth\columnwidth\hsize\csname
@twocolumnfalse\endcsname

\title{
Self-consistent treatment of dynamical correlation functions using a
spectral representation technique}

\author{M. Letz$^{a}$ and F.~Marsiglio$^{b}$\\
$^{a}$ Institut f\"ur Physik, Johannes-Gutenberg Universit\"at, 55099
Mainz, Germany\\ 
$^{b}$ Department of Physics, University of Alberta, Edmonton, Alberta, Canada
T6G 2J1}

\date{\today}

\maketitle

\begin{abstract}
A system of equations resulting from 
an approximation of 
the equation of motion of Green
functions for correlated electron systems is usually solved  using
Matsubara technique. In this work we propose an alternative method
which works entirely along the real frequency axis.
Using the example of the attractive Hubbard model studied in the T-matrix
approximation both self-consistently and non-self-consistently we
demonstrate how powerful such a treatment is especially 
when dynamic quantities are calculated.
\end{abstract}
\pacs{74.20 Mn 74.25.-q 74.20-z}
]
\vskip -0.5cm
%\twocolumn
\narrowtext

\section{Introduction}
\label{sec:intro}
When investigating 
systems in thermodynamic equilibrium with time independent
Hamiltonians 
it was a great success to solve and discuss correlation functions in
the transformed frequency space.
These functions are analytic functions in the complex plane except for
a branch cut along the real axis. This knowledge was first used for the 
T=0 Green function
technique e.g. \cite{zubarev60,tyablikov67}. 
However, especially for finite temperatures, the theory of complex
differentiable functions had led to the development of a very
powerful method, the Matsubara technique \cite{matsubara55}.

For the derivation of the equations of motion an imaginary axis
formulation using Matsubara frequencies was used very early. The
numerical solution of such equations however was first achieved along the
real axis. An example for this is the solution of the Eliashberg
equations by Schrieffer et al. \cite{schrieffer63,scalapino66}.

Later it was discovered that a numerical solution was far more efficient
when an imaginary axis technique is used. In this way static
quantities like e.g. expectation values for occupation numbers $\langle n
\rangle$ or double occupancy $\langle n_{\uparrow} n_{\downarrow}
\rangle $ and their temperature dependence and some thermodynamic
quantities can be calculated successfully. An example is the calculation of
the the superconducting critical temperature \cite{owen71,bergmann73} and the
temperature dependence of the critical magnetic field and specific
heat from the Eliashberg equations \cite{rainer74}.

A major shortcoming of such an imaginary axis treatment is that
dynamical quantities like, for example, the electron density of 
states (along the real
frequency axis) are difficult to obtain. Usually Pade approximants
\cite{vidberg77} or maximum entropy \cite{silver90} techniques are
used to obtain 
dynamical quantities. 
The fact that we need these complicated methods shows how difficult it
is to calculate the values of a function at n$^R$ points  along the
real axis when n$^I$ points along the imaginary axis are known.
The simplest way to illustrate this is the following: The
further away the n$^I$ points are from the n$^R$ points the more the
system of equations which has to be solved in order to obtain
dynamical quantities tends to be singular and therefore tiny inaccuracies
of the functions on the n$^I$ points, the Matsubara frequencies, can
lead to much bigger inaccuracies along the real axis, at points
n$^R$. 

In this paper we 
go in some sense back to the old real axis technique and
propose a treatment in which all
quantities are 
directly calculated along the real axis. 
Since we do not
calculate the functions at the Matsubara frequencies we do not need a
mapping onto the real axis. 
As an example we use the
T-matrix calculation using the ladder diagrams in the
particle-particle channel for the 
attractive Hubbard model. This approximation is valid in the low
density regime and is of particular interest since it might be able to
model some aspects of the short coherence length pairs observed in the
high-T$_c$ superconductors \cite{randeria89,micnas95,janko97}.

\section{spectral representations}
In order to motivate our method we give here a short overview of how
different treatments of Green functions can be connected by using a
spectral representation and by using the whole frequency plane. 

A one particle correlation function of operators $C$ and $B$ 
which are either both Bosonic or Fermionic operators
can be
written as a function of 
temporal and spatial coordinates:
\begin{equation}
\label{eq1:def}
G^k(x_i,t,x_j,t') = -i \langle
T_{F/B} C^{\dagger}(x_i,t) B(x_j,t') \rangle 
\end{equation}
where $T_{F/B}$ is the time ordering operator for Fermions or Bosons and 
\begin{equation}
\langle ... \rangle =
Z^{-1} Tr(e^{-\beta \tilde{H}}...)
\end{equation}
is the thermal expectation value. $Z$ is the partition function in the
grand canonical ensemble and $ 
\tilde{H} = H - \mu N $ is the Hamiltonian in this ensemble. 
If we restrict ourselves to systems 
in thermal equilibrium
with periodic boundary conditions and
time independent Hamiltonians we can apply a
Fourier transform \cite{laplace}:
\begin{eqnarray}
\label{eq2:def}
G^k({\bf k},\omega) &=& \int_{-\infty}^{\infty} d(t-t') \frac{1}{N} 
\int d({\bf x}_i - {\bf x}_j) \nonumber \\ && G^k(t-t',x_i-x_j) 
e^{i \omega (t-t')} e^{i {\bf k}({\bf x}_i - {\bf x}_j)} \;\;\; .
\end{eqnarray} 
The definition (\ref{eq1:def}) and equation (\ref{eq2:def}) are valid at
all temperatures.\\ 
The function  $G^k({\bf k},\omega)$ is related to a function which is
analytic 
in the whole complex $\omega$-plane with the exception of a branch
cut along the real axis. Therefore there exists a spectral
representation and $G^k({\bf k},\omega)$ can be rewritten as a
function of a spectral function $J({\bf k},\omega)$ which is a
real function defined along the real $\omega$ axis. The Green function
$G^k({\bf k},\omega)$ is connected to its spectral representation in
the following way:
\begin{eqnarray}
\label{eq:spect}
G^k({\bf k},\omega) &=& \lim_{\delta \rightarrow 0^+} \frac{1}{2\pi} 
\int_{-\infty}^{\infty} d\overline{\omega} \nonumber \\ &&
\biggl( \frac{J({\bf k},\overline{\omega}) e^{\beta \overline{\omega}}}
{\omega -   \overline{\omega} + i \delta} - 
\frac{\mp J({\bf k},\overline{\omega}) }
{\omega -   \overline{\omega} - i \delta}  \biggr) \;\;\; .
\end{eqnarray} 
Here and in the following the upper sign
corresponds to Fermions whereas the lower sign describes Bosons. For
practical use we want to define a slightly different function $A({\bf
k},\omega)$:
\begin{equation}
A({\bf k},\omega) = \frac{1}{2} J({\bf k},\omega) (e^{\beta
\omega} \pm 1) \;\;\; .
\end{equation}
Therefore equation (\ref{eq:spect}) can be rewritten:
\begin{eqnarray}
G^k({\bf k},\omega) &=& \lim_{\delta \rightarrow 0^+}
\frac{1}{\pi}
\int_{-\infty}^{\infty} d\overline{\omega} \nonumber \\ &&
\biggl(
\frac{A({\bf k},\overline{\omega}) 
\frac{e^{\beta \overline{\omega}}}{e^{\beta \overline{\omega}} \pm 1}
}
{\omega -   \overline{\omega} + i \delta} \pm 
\frac{ A({\bf k},\overline{\omega}) 
\frac{1}{e^{\beta \overline{\omega}} \pm 1}
}
{\omega -   \overline{\omega} - i \delta} \biggr) \;\;\; .
\end{eqnarray} 
For a non-interacting system and $C$,$B$ e.g. Fermionic operators
$c_{\bf k}$ the function $\sum_{\bf k} A({\bf k},\omega)$ is identical to the
density of states. For an interacting system $\sum_{\bf k} A({\bf k},\omega)$
will still be a density of states but $A({\bf k},\omega)$ becomes dependent on the
thermodynamic variables $T$,$\mu$.
\begin{equation}
A({\bf k},\omega) \longrightarrow A^{T,\mu}
({\bf k},\omega) 
\end{equation}
The usual $T=0$ Green functions for which the zero temperature diagram
technique \cite{fetwal} is valid can be rewritten as a function of 
$A^{T,\mu}
({\bf k},\omega)$:
\begin{eqnarray}
G^k({\bf k},\omega) &=& \lim_{\beta \rightarrow \infty} 
\lim_{\delta \rightarrow 0^+}
\frac{1}{\pi}
\int_{-\infty}^{\infty} d\overline{\omega} \nonumber \\ &&
\biggl(
\frac{A^{T,\mu}({\bf k},\overline{\omega}) 
\frac{e^{\beta \overline{\omega}}}{e^{\beta \overline{\omega}} \pm 1}
}
{\omega -   \overline{\omega} + i \delta} \pm 
\frac{ A^{T,\mu}({\bf k},\overline{\omega}) 
\frac{1}{e^{\beta \overline{\omega}} \pm 1}
}
{\omega -   \overline{\omega} - i \delta} \biggr) \\
G^R({\bf k},\omega) &=& \lim_{\beta \rightarrow \infty} 
\lim_{\delta \rightarrow 0^+} 
\frac{1}{\pi}
\int_{-\infty}^{\infty} d\overline{\omega}
\frac{A^{T,\mu}({\bf k},\overline{\omega})}
{\omega -   \overline{\omega} + i \delta} \\
G^A({\bf k},\omega) &=& \lim_{\beta \rightarrow \infty} 
\lim_{\delta \rightarrow 0^+} 
\frac{1}{\pi}
\int_{-\infty}^{\infty} d\overline{\omega}
\frac{A^{T,\mu}({\bf k},\overline{\omega})}
{\omega -   \overline{\omega} - i \delta} 
\end{eqnarray} 
where $R,A$ denotes retarded and advanced Green functions. $G^R({\bf 
k},\omega)$ and $G^A({\bf k},\omega)$ are both branches of one
function $G({\bf k},z)$
defined on the whole complex plane with the exception of the
branch cut along the real axis where the poles of $A^{T,\mu}({\bf
k},\overline{\omega})$ are located:
\begin{equation}
G({\bf k},z) = 
\frac{1}{\pi}
\int_{-\infty}^{\infty} d\overline{\omega}
\frac{A^{T,\mu}({\bf k},\overline{\omega})}
{z - \overline{\omega} } \;\; .
\end{equation}

Also the thermal Green functions defined entirely along the
imaginary frequency axis can be written as a function of 
$A^{T,\mu}({\bf k},\overline{\omega})$ 
\begin{equation}
\label{eq:therm}
G^t({\bf k},i\omega_n) = 
\frac{1}{\pi}
\int_{-\infty}^{\infty} d\overline{\omega}
\frac{A^{T,\mu}({\bf k},\overline{\omega})}
{i\omega_n -\overline{\omega}  }
\end{equation}
where $i \omega _n = \frac{(2n+1)\pi i}{\beta} , \frac{2n\pi i}{\beta} $
are the Matsubara frequencies for
Fermions and Bosons respectively. The fact that the functions are
only defined at certain periodic points means that the Fourier
series of eq. (\ref{eq:therm}) 
\begin{equation}
G^t({\bf k},i\tau) = \frac{1}{\beta} \sum_{n=-\infty}^{\infty}
e^{-i\omega _n \tau} 
G^t({\bf k},i\omega_n)
\end{equation}
is periodic in imaginary time
\begin{equation}
G^t({\bf k},i\tau + i \beta) = \mp  G^t({\bf k},i\tau) \;\;\; .
\end{equation}
This means that the full knowledge of the function $G({\bf k},z)$ is
either obtained by knowing (a) $G({\bf k},i\omega_n)$ on the infinite
but discrete points $i \omega_n$ along the imaginary axis
or by knowing (b) $A^{T,\mu}({\bf k},\omega)$ on a continuum
along the real axis. The details of the usual Green function technique are
outlined in many textbooks e.g. \cite{AGD,fetwal}. Here we only want
to highlight the connection of both methods with the function 
 $A^{T,\mu}({\bf k},\omega)$. The usual Matsubara technique determines 
the functions along the points $i\omega_n$ whereas we will discuss a
method in this paper which calculates an approximation of 
$A^{T,\mu}({\bf k},\omega)$.

\section{numerical technique}
The solution of the equations in a certain approximation for a
correlated quantum system will generally be found numerically.
In order to achieve self-consistency it is
preferable to
perform discrete sums over Matsubara frequencies than to calculate a
function like  $A^{T,\mu}({\bf k},\omega)$. But to obtain dynamical
quantities along the real axis, a difficult and somewhat uncontrolled
analytic continuation will have to be performed.
Therefore there have been attempts to solve such systems along the
real axis (see for example  \cite{marsiglio88} for the Eliashberg
equations and \cite{fresard92,kyung98} for the self-consistent T-matrix
equations). In these works some numerical integration  was required
along the real axis.

However in this section we
argue 
that it is possible to replace $A^{T,\mu}({\bf k},\omega)$ by a series
of (typically a few hundred) $\delta$ functions along the real
axis. With this approximation all frequency integrations will turn
into summations over a finite number of $\delta$ functions and can
therefore be done analytically. We use
\begin{equation}
A^{T,\mu}({\bf k},\omega)
\approx \pi \sum_{l=1}^{N_{max}} a_l^{\bf k} \delta(\omega - b_l)
\end{equation}
where the amplitudes are $a_l^{\bf k}$ (which also
depend 
on the thermodynamic variables)
and the poles are located at
the positions $b_l$. The Green function $G({\bf k},\omega)$ can now be
expressed in this approximation as a sum of poles:
\begin{eqnarray}
 G({\bf k},i\omega_n) & \approx &
\frac{1}{\pi}
\int_{-\infty}^{\infty} d\overline{\omega}
\frac{\pi \sum_{l=1}^{N_{max}} a_l^{\bf k} \delta(\overline{\omega} - b_l)}
{i\omega_n -\overline{\omega}  }
\nonumber \\ &=&
\sum_{l=1}^{N_{max}} \frac{a_l^{\bf k}}{i\omega_n - b_l} 
\;\;\; .
\end{eqnarray} 
Using such a spectral representation our aim is to convert the usually
complicated equations for $G({\bf k},i\omega_n)$ into sets of
equations for the amplitudes $a_l^{\bf k}$ only.

\subsection{Frequency grid}
An important point is how to choose the frequency points $b_l$ along the
real axis. In a previous work \cite{letz98} 
we let them
fluctuate freely during the calculation but employed some
approximations to restrict the number $N_{max}$. In this work we keep
them fixed relative to the chemical potential
which leads to an efficient algorithm.\\
It also turned out to be of importance to adopt the frequency grid to
the problem (e.g. if the influence of a band edge is important the
frequency points should be more dense around that band edge)
and especially to the temperature.\\
For a Fermionic system -- where the energy range of $\pm k_BT$ around the
chemical potential is of importance -- we choose for the example
discussed in section \ref{sec:neguhub}:
\begin{equation}
\label{eq:grid1}
b_l = \frac{N_{max}-1}{\beta \; \alpha} \tanh ^{-1} \left (
\frac{ N_{max}-l}{N_{max}-1} h_1 + 
\frac{ l-1}{N_{max}-1} h_2  \right )
\end{equation}
with
\begin{equation}
\label{eq:grid2}
h_{1/2} = \tanh \left (
\frac{\beta \; \alpha \; \omega_{min/max}}
{N_{max}-1} \right )
\end{equation}
and $\omega_{min}$ and $\omega_{max}$ are the minimum and maximum frequencies
considered, respectively. 
The parameter $\alpha$ was adjusted in the way that the
distance between two frequency points around the chemical potential is
always smaller than $k_BT$. In this way the delta functions are
arranged with highest density at the chemical potential and far away
from the chemical potential they get thinned out. Note that the index
$l$ does not need to have integer values -- this will be used later on.

\subsection{Products of correlation functions}
\label{sub:susz}
Usually in such calculations we have to deal with products of
correlation functions which can be folded both in momentum and
frequency space. 
The result of such products of one-particle
correlation functions are generalized susceptibilities. If we have two
functions $G^1({\bf k},i\omega_n)$, $G^2({\bf k},i\omega_n)$, we
often need to calculate the following product:
\begin{eqnarray}
\chi({\bf K},i\Omega_m) & = & -\frac{1}{\beta} \sum_n \frac{1}{N} \sum_{\bf q}
\nonumber \\  & & \;\;\;\;
G^1({\bf q},i\omega_n) \; G^2({\bf K-q},i \Omega_m - i\omega_n)
\end{eqnarray}
where ${\bf K} $ is the total momentum of a pair of particles.
We denote in the following Bosonic Matsubara
frequencies and q-vectors of Bosons with upper-case letters whereas
for Fermionic systems we use lower-case letters.
$G^{i}({\bf k},i\omega_n)$, for $i=1,2$, has the approximate spectral
representation: 
\begin{equation}
G^{i}({\bf q},i\omega_n) \approx \sum_{j=1}^{N_{max}} \frac{a^{i \;
{\bf k} }_j}{ i\omega_n - b_j} \;\; .
\end{equation}
We can directly evaluate the frequency summations and are left with
sums over the coefficients.
When inserting the spectral representation for $G^{i}({\bf k}, i \omega _n)$ 
we get:
\begin{eqnarray}
\label{chispr}
\lefteqn{
\chi({\bf K}, i \Omega _n) = - \sum_{\bf q} \sum_{j,l}^{N} \frac{1}{\beta} \sum_
m 
\frac{a_l^{1 \; {\bf q}}}{i \Omega _n - i \omega_m - b_l} \;
\frac{a_j^{2 \; {\bf K}-{\bf q}}}{i \omega _m - b_j}} \nonumber \\
&=&
\sum_{\bf q} \sum_{j,l}^{N}  
\frac{ a_l^{1 \; {\bf q}} \; a_j^{2 \; {\bf K}-{\bf q}}} {i \Omega _n -
b_j - b_l} \left  
(
\frac{1}{1+e^{\beta b_j}} - \frac{1}{1+e^{- \beta b_l}} \right )  
\nonumber \\ &=&
\frac{1}{2} \sum_{\bf q} \sum_{j,l}^{N} \frac{a_l^{1 \; {\bf 
q}} \; a_j^{2 \; {\bf K}-{\bf q}} }{i \Omega _n - b_j -
b_l} 
\nonumber \\ && \;\;\;\;\;\;\;\;\;\;\;\;\;\;\;\;
\left ( \tanh \left ( \frac{\beta b_j}{2} 
\right ) + \tanh \left ( \frac{\beta b_l}{2} 
\right ) \right )  
\end{eqnarray} 
We now have determined a spectral representation of $\chi({\bf K},Z)$
which is valid in the whole complex plane. Nonetheless, for convenience we
continue writing our function on the Matsubara frequencies.
$\chi({\bf K}, i \Omega _n)$ is however defined via a spectral
representation 
on $N_{max}(N_{max}+1)/2$ frequency points
along the real frequency axis which have to be folded back onto the
$N_{max}$ points of our frequency grid.
Therefore we sort the $N_{max}(N_{max}+1)/2$ points
\begin{equation}
\label{sort}
\tilde{b}^{\bf K}_{jl} = b_j + b_l \;\;\; ,
\end{equation}
check if they fit into each interval $b_{p-1/2} < \tilde{b}^{\bf
K}_{jl} < b_{p+1/2} $
(the index of b in eq. (\ref{eq:grid1}) can be non integer) 
and add their amplitudes
\begin{equation}
\tilde{c}^{\bf K}_{jl} =  
a_l^{1 \; {\bf 
q}} \; a_j^{2 \; {\bf K}-{\bf q}}
\left ( \tanh \left ( \frac{\beta b_j}{2} 
\right ) + \tanh \left ( \frac{\beta b_l}{2} 
\right ) \right )  
\end{equation}
to the amplitude $c^{\bf K}_{p}$ of the susceptibility. In this way we
obtain a similar spectral representation for the susceptibility:
\begin{equation}
\chi({\bf K}, i \Omega _n) = \sum_{p=1}^{N_{max}} 
\frac{c^{\bf K}_{p}}{i \Omega_n - b_p} \;\; .
\end{equation}
For the further calculation we just have to store the amplitudes $c^{\bf
K}_{p}$. Note that if the $G^{i}({\bf q},i\omega_n), i = 1,2$ had 
been Fermionic
correlation functions, the amplitudes $c^{\bf
K}_{p}$ are -- as opposed to the amplitudes $a^{i \;
{\bf k} }_j$ -- not only positive but change sign at the chemical
potential due to the $\tanh$ function. The function $\chi({\bf
K},i\Omega_m)$ will therefore be a Bosonic correlation function and
$i\Omega_m$ are Bosonic Matsubara frequencies.

\subsection{Vertex functions}

Vertex functions $\Gamma({\bf k}_1 , {\bf k}_2, ... , i \omega _{n1}, i
\omega _{n2}, ...)$
result from Bethe-Salpeter type equations and they
are in general functionals of two particle propagators
similar to the one described in subsection \ref{sub:susz}, 
of a
correlation $U_{corr}$ (which in general is a function of ${\bf k}$ and
$\omega$ as well)
and of one particle correlation functions.
\begin{eqnarray}
\lefteqn{
\Gamma({\bf k}_1 , {\bf k}_2, i \omega _{n1}, i
\omega _{n2}) = \nonumber} \\
&&\Gamma\{\chi({\bf k}_1 + {\bf k}_2 , i \omega _{n1} + i \omega _{n1}),
... , 
U_{corr}, G ({\bf k}_i, i \omega _{ni}), ...   
\} \;\; .
\end{eqnarray}
If they can be reduced to analytic functions in the complex plane they
can be calculated by shifting the poles of the spectral representation
at $b_n$ 
into the upper half plane by an amount
\begin{equation}
\delta_n = \frac{1}{2} (b_{n+1/2} - b_{n-1/2}),
\end{equation}
where the shift into the upper half plane depends on $n$
and is smallest close to the chemical potential.
With the help of such a vertex function a proper self-energy can be
calculated in a way similar to the calculation of the susceptibility
in subsection \ref{sub:susz}.
In subsection \ref{sub:vertnegu} this is shown on the example of the
attractive Hubbard model in the ladder approximation.

\section{example: Attractive Hubbard model in the ladder approximation}
\label{sec:neguhub}
The attractive Hubbard model shows superconductivity in its weak
coupling 3D limit. The range of intermediate coupling and 2-3 dimensions
is of particular interest for the high-T$_c$ cuprates
\cite{micnas95}. The Hubbard Hamiltonian is
\begin{equation}
\label{eq:hub}
H = -t \sum_{<i,j>,\sigma} c^{\dagger}_{i\sigma} c_{j\sigma} + 
U \sum_{i}  n_{i\uparrow} n_{j\downarrow},
\end{equation}
where $t$ is the transfer integral between two neighbouring lattice sites
$<i,j>$ (we restrict ourselves to hyper-cubic lattices) and $U<0$ is the
attractive interaction for two electrons on 
the same site of the lattice. In the case of $U=0$ the Hamiltonian can
be diagonalized and gives the usual dispersion of a tight binding
Hamiltonian $\epsilon({\bf k})$.\\
The non-interacting Green function $G^{(0)}({\bf k}, i \omega _n)$
is approximated by the pole $b_j$ which is nearest to
$\epsilon({\bf k})$.
\begin{equation}
G^{(0)}({\bf k}, i \omega _n) \approx \frac{1}{i \omega _n
-b_j}
\end{equation}   
\begin{equation}
b_{j-1/2} < \epsilon({\bf k}) \le b_{j+1/2} \;\;\;\;\; .
\end{equation}
The ladder
approximation \cite{fetwal} takes into account all non-crossing
scattering events of a pair in the particle-particle channel and becomes
exact for all coupling strengths in the low density limit $n
\longrightarrow 0$.  

\begin{figure}
\unitlength1cm
\epsfxsize=9cm
\begin{picture}(7,3.5)
\put(-1.0,-5.5){\epsffile{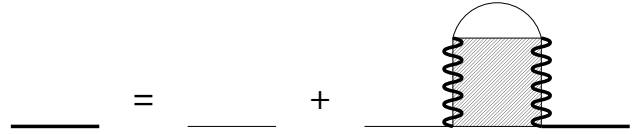}}
\end{picture}
\caption{ 
schematic diagram of the non-self-consistent ladder equations. The
full thick lines represent full, interacting Green functions, the thin
lines non interacting Green functions and the wavy lines are the
Hubbard interaction U. 
}
\label{fignsc}
\end{figure}

This leads to the following system of equations \cite{haussmann93} which has to
be solved either non self-consistently (Eqs. (\ref{nsc1})-(\ref{nsc4}))
which is diagrammatically shown in Fig. \ref{fignsc}
\begin{eqnarray}
G^{(0)}({\bf k}, i \omega _n) &=& \frac{1}{i \omega _n - \epsilon({\bf
k})} \\
\label{nsc1}
\chi^{(0)}({\bf K}, i \Omega _m) &=& 
-\frac{1}{\beta} \sum_n \frac{1}{N} \sum_{\bf q}
\nonumber \\  & & \mbox{\hspace*{-0.5cm}}
G^{(0)}({\bf q},i\omega_n) \; G^{(0)}({\bf K-q},i \Omega_m - i\omega_n) \\
\label{gam:nsc}
\tilde{\Gamma}^{(0)}({\bf K}, i \Omega _n) &=&
\frac{U^2 \chi^{(0)}({\bf K}, i \Omega _n)}{
1-U \chi^{(0)}({\bf K}, i \Omega _n)} + U \\
\Sigma^{(0)}({\bf k}, i \omega _n) &=& 
\frac{1}{\beta} \sum_m \frac{1}{N} \sum_{\bf q}
\nonumber \\  & & \mbox{\hspace*{-0.5cm}}
\tilde{\Gamma}^{(0)}({\bf k} + {\bf q}, i \omega _n + i \omega _m)
G^{(0)}({\bf q},i\omega_m)\\
\label{nsc4}
G^{nsc}({\bf k}, i \omega _n) &=& \frac{1}{G^{(0)}({\bf k}, i \omega
_n)^{-1} - \Sigma^{(0)}({\bf k}, i \omega _n)}
\end{eqnarray}
\begin{figure}
\unitlength1cm
\epsfxsize=9cm
\begin{picture}(7,3.5)
\put(-1.0,-5.5){\epsffile{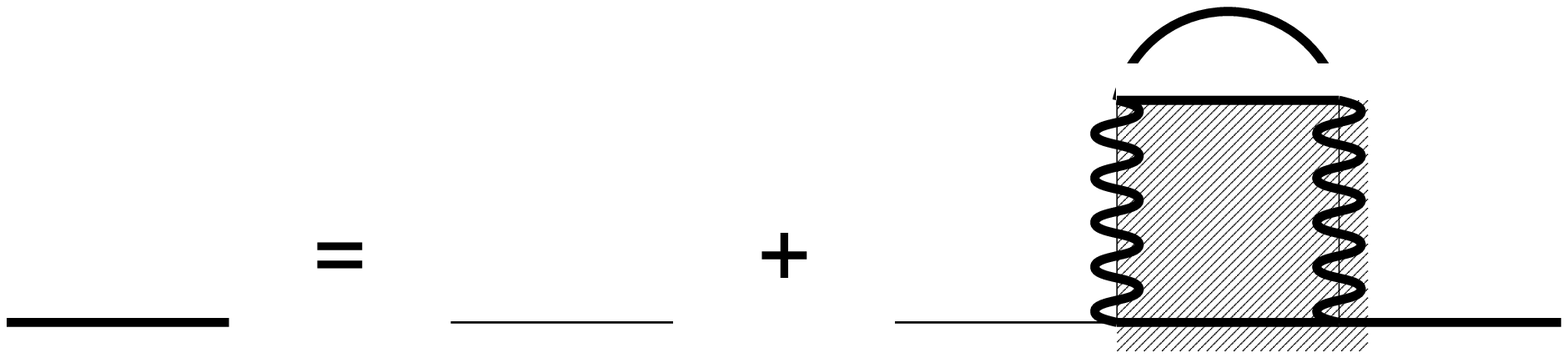}}
\end{picture}
\caption{ 
schematic diagram of the fully self-consistent ladder equations.
}
\label{figsc}
\end{figure}
or self-consistently (Eqs. (\ref{sc1})-(\ref{sc4})) 
which is diagrammatically shown in Fig. \ref{figsc}
\begin{eqnarray}
\label{sc1}
\chi({\bf K}, i \Omega _m) &=& 
-\frac{1}{\beta} \sum_n \frac{1}{N} \sum_{\bf q}
\nonumber \\  & & \;\;\;\;
G({\bf q},i\omega_n) \; G({\bf K-q},i \Omega_m - i\omega_n) \\
\label{gam:scf}
\tilde{\Gamma}({\bf K}, i \Omega _n) &=&
\frac{U^2 \chi({\bf K}, i \Omega _n)}{
1-U \chi({\bf K}, i \Omega _n)} + U \\
\Sigma({\bf k}, i \omega _n) &=& 
\frac{1}{\beta} \sum_m \frac{1}{N} \sum_{\bf q}
\nonumber \\  & & \;\;\;\;
\tilde{\Gamma}({\bf k} + {\bf q}, i \omega _n + i \omega _m)
G({\bf q},i\omega_m)\\
\label{sc4}
G({\bf k}, i \omega _n) &=& \frac{1}{G^{(0)}({\bf k}, i \omega
_n)^{-1} - \Sigma({\bf k}, i \omega _n)} \;\;\; .
\end{eqnarray}
In order to obtain self-consistency the Eqs. (\ref{sc1})-(\ref{sc4})
have to be calculated iteratively until a stable self-consistent
solution is obtained.

\subsection{Vertex function in the ladder approximation }
\label{sub:vertnegu}
In the case of the ladder approximation the vertex function
(Eqs. (\ref{gam:nsc}),(\ref{gam:scf})) itself 
does not tend to zero as $\Omega \longrightarrow
\infty$, instead $\lim_{\Omega \longrightarrow
\infty} = U $. However the function $\tilde{\Gamma}({\bf
K},i\Omega_m)$ can be decomposed into a function which fulfills
Kramers-Kronig relations 
$\Gamma({\bf K},i\Omega_m)$ plus a constant $U$ \cite{thoulesskrit}.
The function  $\Gamma({\bf K},i\Omega_m)$ is $U^2$ times the 
two-particle propagator of a pair with total momentum ${\bf K}$.
In order to obtain a spectral representation for $\Gamma({\bf K},z)$ we
have two possibilities, which we now discuss in turn.
\subsubsection{Complex evaluation}
\label{suba}
In the following we show the
evaluation of $\Gamma({\bf K},z)$ in the complex plane. Therefore we
need an approximate expression for the complex function $\chi({\bf
K},z)$ at the frequency 
points $b_m$ of our frequency grid of the real axis.
\begin{equation}
\chi({\bf K},b_m) \approx \sum_{n=1}^{N_{max}} 
\frac{c_m^{\bf K}}{b_m - b_n + i\delta_n }
\end{equation}
with
\begin{equation}
\label{eps}
\delta_n = \frac{1}{2} (b_{n+1/2}-b_{n-1/2})
\end{equation}
This has to be put into the equation for $\Gamma$
\begin{equation}
\label{gam:ot}
\Gamma({\bf K},b_m) \approx \frac{ U^2 \chi({\bf K},b_m)}{
1-U \;  \chi({\bf K},b_m)}
\end{equation}
The amplitudes $g_m^{\bf K}$ of $\Gamma({\bf K},b_m)$ at the points $b_m$
are then given by:
\begin{equation}
g_m^{\bf K} \approx \frac{2 \delta_m}{\pi} \Im(\Gamma({\bf K},b_m))
\end{equation}
and we have obtained a spectral representation for $\Gamma({\bf K},b_m)$
which can be used for further calculation:
\begin{equation}
\Gamma({\bf K},i \Omega_m) \approx \sum_{n=1}^{N_{max}} \frac{
g_m^{\bf K}}{i \Omega_m - b_n} 
\end{equation} 
\subsubsection{ Evaluation with partial fractions}
\label{subb}
A second possibility to calculate the amplitudes $g_m^{\bf K}$ is by
rewriting Eq. (\ref{gam:ot})
\begin{eqnarray}
\label{partfrac}
\lefteqn{\Gamma({\bf K},\Omega) =}\\
&& \frac{U^2 \;\sum_{n=1}^{N_{max}}  
\prod_{{m=1}\atop {m \neq n}}^{N_{max}}  
(\Omega - b_m) \; c_n^{\bf K} }
{\prod_{n=1}^{N_{max}} (\Omega - b_n) - U 
\sum_{n=1}^{N_{max}}  
\prod_{{m=1}\atop {m \neq n}}^{N_{max}}  
(\Omega - b_m) \; c_n^{\bf K}}
\end{eqnarray}
and looking for the poles of the denominator. Since it is a polynomial
of order $N_{max}$ this seems to be a hopeless procedure. However the
roots are bracketed between the old poles $b_m$ and we know that
the number of roots between $b^{\bf K}_m$ and $b^{\bf K}_{m+1}$ 
is one except at zero frequency were it can not exceed two.
Therefore this problem can be solved numerically. Having found the
roots $\tilde{b}^{\bf K}_m$ we can evaluate the amplitudes by putting
the solution into the numerator of Eq. (\ref{partfrac}). This
procedure is just calculating partial fractions for eq. 
(\ref{partfrac}). An analogous procedure as for the susceptibility
(see Eq. (\ref{sort})) allows us to calculate the amplitudes $g^{\bf
K}_m$ at the frequency points $b_m$ by adding all amplitudes for
frequencies $\tilde{b}^{\bf K}_m$ in the interval $b_{m-1/2} < 
\tilde{b}^{\bf K}_m < b_{m+1/2}$.\\[0.2cm]
Both methods \ref{suba} and \ref{subb} work well. In \cite{letz98}
we used method \ref{subb} and for the calculations presented in this
paper  we use method \ref{suba}.

\subsection{Self energy}
Having calculated the vertex function $\Gamma({\bf K},i \Omega_m)$ we
can proceed and calculate the self-energy $\Sigma'({\bf k},i \omega_n)$.
The ' just reminds us that we do not calculate the full self-energy
since we use $\Gamma({\bf K},i \Omega_m)$ instead of
$\tilde{\Gamma}({\bf K},i \Omega_m)$. The inclusion of the frequency
independent part ($U$) of the vertex function will lead to the Hartree
term. For the frequency dependent self-energy, we have
\begin{eqnarray}
\lefteqn{\Sigma'({\bf k},i \omega_n) =} \nonumber \\
&&\frac{1}{N} \sum_{\bf q} \frac{1}{\beta} \sum_m 
\Gamma({\bf k} +{\bf q}  ,i \omega_n + i \omega_m) \;
G({\bf q},i \omega_m) \nonumber \\
&\approx &    
 \frac{1}{N} \sum_{\bf q} \sum_{j,l}^{N_{max}} \frac{1}{\beta} \sum_m
\frac{g^{{\bf k} +{\bf q}}_l }
{i \omega_n + i \omega_m - b_l} \;
\frac{a^{\bf q}_j}{i \omega_m - b_j} \nonumber \\
&=&
\frac{1}{N} \sum_{\bf q} \sum_{j,l}^{N_{max}}
\frac{g^{{\bf k} +{\bf q}}_l \; a^{\bf q}_j}{i \omega _n - b_l + b_j} 
\left (
\frac{1}{1+e^{\beta b_j}} + \frac{1}{e^{\beta b_j}-1} 
\right )
\end{eqnarray}
The Bosonic distribution function is due to the Bosonic nature of a
pair of Fermions described by $\Gamma({\bf K}  ,i \Omega_m)$. Again an
analogous procedure as for the susceptibility (see Eq. (\ref{sort}))
yields the coefficients $s^{\bf k}_j$ at the frequencies $b_j$ of our
frequency grid. In this way we obtain a spectral representation for
the self-energy
\begin{equation}
\Sigma'({\bf k},i \omega_n) = \sum_{n=1}^{N_{max}} \frac{s^{\bf k}}
{i \omega_n - b_j} \;\;\; .
\end{equation}
\subsection{Calculation of the full Green function}
Knowing the self-energy $\Sigma({\bf k},i \omega_n)$ we can calculate
the Green function for the one-particle propagator.
\begin{eqnarray}
G({\bf k},i \omega_n) &=& \left (
i \omega_n - \epsilon({\bf k}) + \mu - \Sigma({\bf k},i \omega_n)
\right )^{-1} \nonumber \\
&\approx & \left (
i \omega_n - \epsilon({\bf k}) + \mu - \sum_{j=1}^{N_{max}}
\frac{s^{\bf k}}  {i \omega_n - b_j}
\right )^{-1} 
\end{eqnarray}
Again we have two alternatives to calculate a spectral representation
for $G({\bf k},i \omega_n)$. Analogous to the case \ref{suba} when
calculating the vertex function from the 
susceptibility we get for the amplitudes of the one-particle Green
function,  
\begin{eqnarray}
a^{\bf k}_m &=& \frac{2 \delta _m }{\pi} \Im ( G({\bf k},b_m))
\nonumber \\ & \approx & 
\label{procend}
\frac{2 \delta _m }{\pi} \Im \left (
b_m  - \epsilon({\bf k}) + \mu - \sum_{j=1}^{N_{max}}
\frac{s^{\bf k}_n}{b_m - b_j + i \delta _ j} \right )^{-1}
\end{eqnarray}
where $\delta_n$ is defined in Eq. (\ref{eps}).
Due to the approximations made the sum rules might not always be
fulfilled. Therefore it is necessary to perform a sum rule check and
correct (if needed) the amplitudes $a^{\bf k}_m$.

\subsection{Self-consistent vs. non-self-consistent} 
Going once through the procedure Eqs. (\ref{nsc1})-(\ref{nsc4})
we obtain a  non-self-consistent result which is conserving in a
one-particle picture (the equation $0<N<2$  with $N$ the total particle
number is fulfilled for all possible 
values of $\mu$ and $T$). Already this is an improvement to the
solution of the T-matrix equations given by Schmitt-Rink et al. \cite{svr},
as advocated by Serene \cite{serene89}.

It is also possible to go iteratively through the
Eqs. (\ref{sc1})-(\ref{sc4}). In this case the amplitudes will need an
additional index $p$ for the number of iterations. 
\begin{equation}
a^{\bf k}_j \longrightarrow a^{{\bf k},p}_j \;\;\; , \;\;\;
c^{\bf K}_j \longrightarrow c^{{\bf K},p}_j \;\;\; , \;\;\; ...
\end{equation}
Now the Eqs. (\ref{sc1})-(\ref{sc4}) just map the amplitudes to the
next level of iteration.
\begin{equation}
a^{{\bf k},p}_j \longrightarrow a^{{\bf k},p+1}_j \;\;\; , \;\;\;
c^{{\bf K},p}_j \longrightarrow c^{{\bf K},p+1}_j \;\;\; , \;\;\; ...
\end{equation}
This procedure has to be repeated until a stable self-consistent
solution is achieved. As a condition for self-consistency we used:
\begin{equation}
\frac{1}{N_{max}} \frac{1}{N_{\bf K}} \left (
\sum_{\bf k} \sum_{j=1}^{N_{max}} ( (a^{{\bf k},p}_j)^2 -
(a^{{\bf k},p+1}_j)^2)\right ) ^{1/2} < \delta
\end{equation}
were $\delta$ was typically chosen around $10^{-7}$.
\section{numerical discussion of spectral quantities}
In the following we will present several dynamical quantities which
were obtained by using a numerical solution of the
equations of the T-matrix in the ladder approximation
for the attractive Hubbard model on a 2D square lattice.

\begin{figure}
\unitlength1cm
\epsfxsize=10cm
\begin{picture}(7,16)
\put(-2.2,7){\rotate[r]{\epsffile{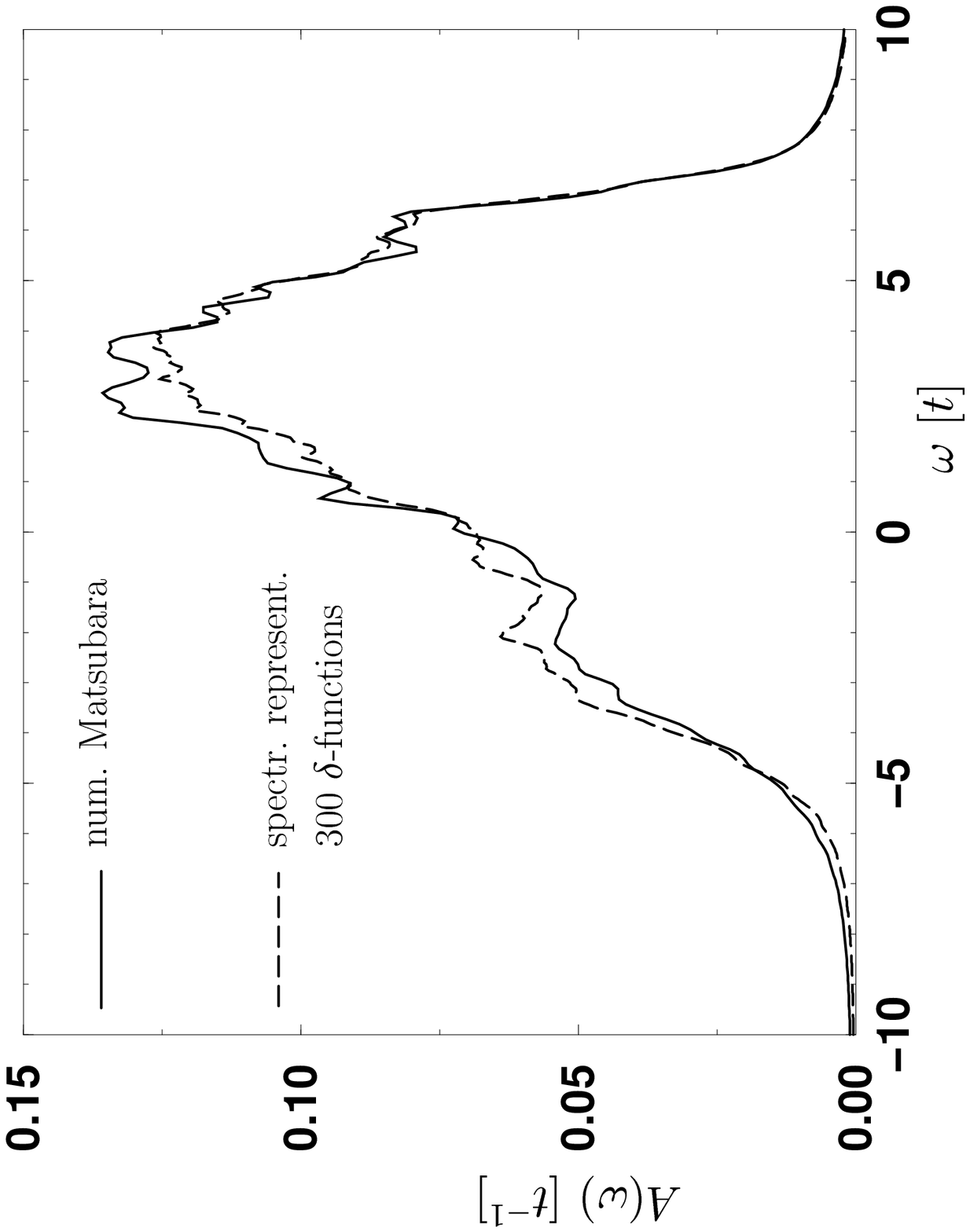}}}
\epsfxsize=10cm
\put(-2.2,0){\rotate[r]{\epsffile{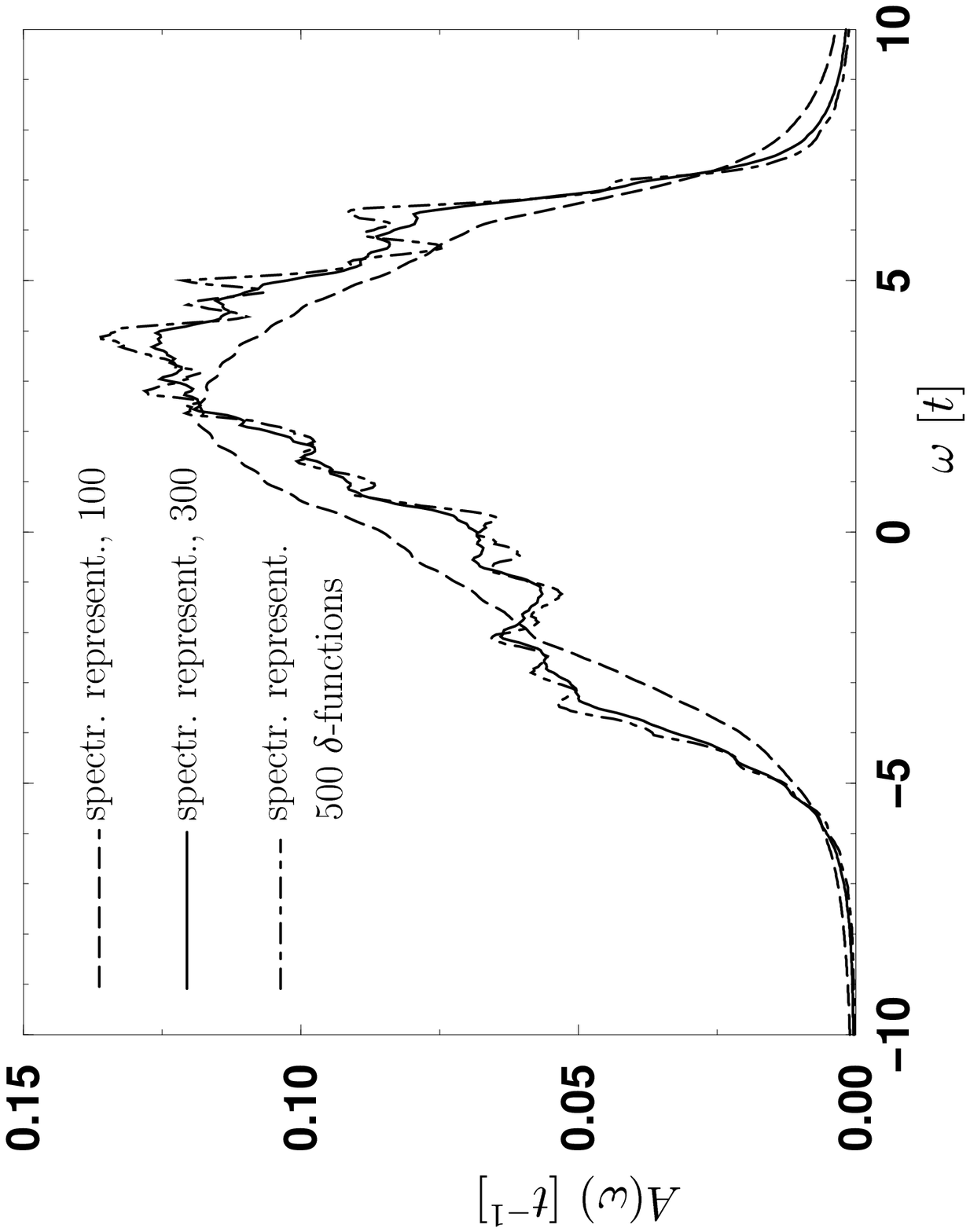}}}
\end{picture}
\caption{For a 8x8 cubic 2D lattice, $k_BT=0.55[t]$, $U=-4[t]$,
$\mu=-2[t]$ and $n \approx 0.7$ the density of states is plotted. 
$\omega=0$ is the position of the chemical potential.
In (a) the
full line corresponds to results obtained with Matsubara technique and
Pade approximates and the dashed line is the result of our spectral
representation technique for $N_{max}=300$. In (b) we compare different
numbers of $N_{max} = 100, 300, 500$. 
}
\label{fig:1}
\end{figure}

\begin{figure}
\unitlength1cm
\epsfxsize=7.5cm
\begin{picture}(7,23)
\put(-1.2,16.5){\rotate[r]{\epsffile{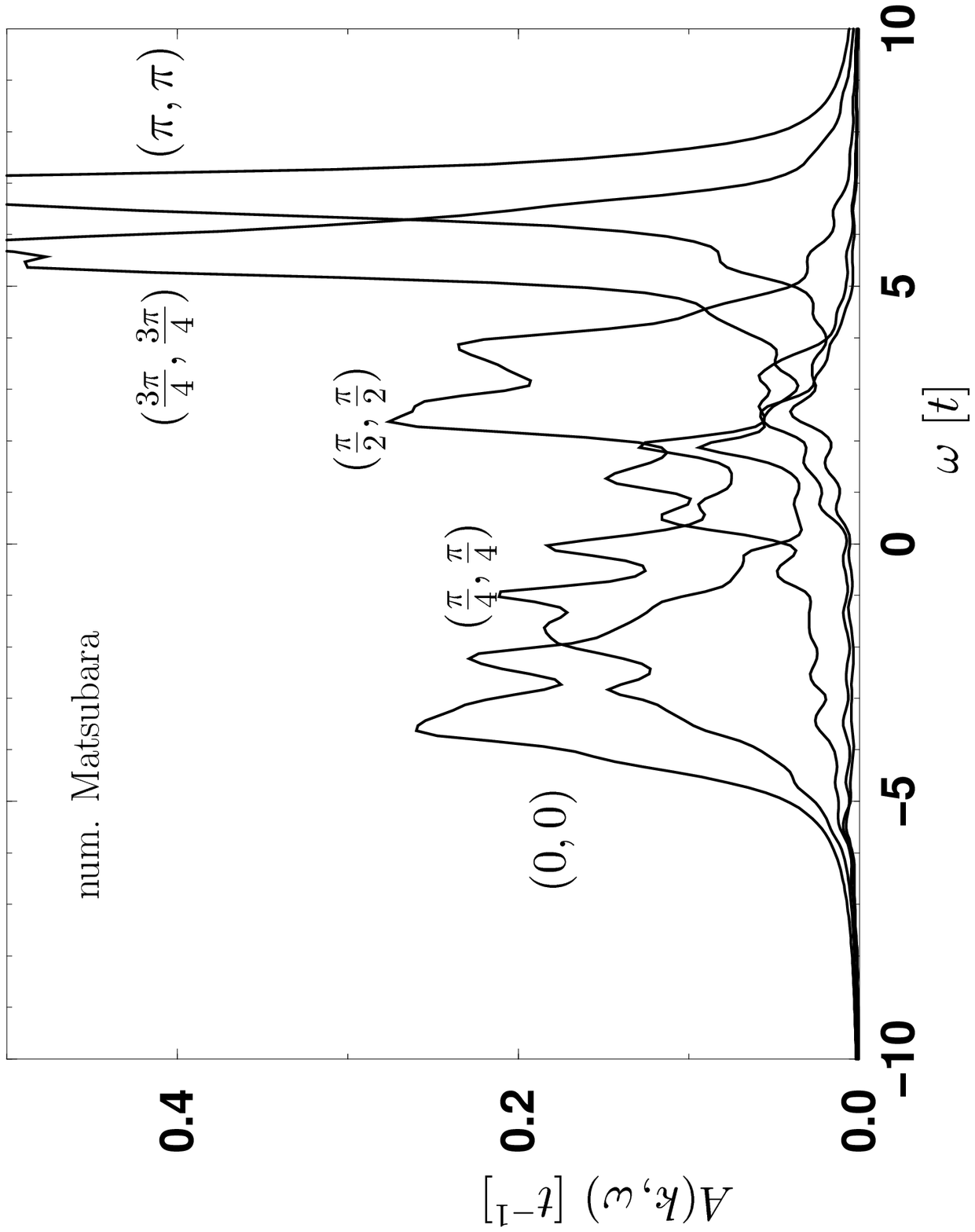}}}
\epsfxsize=7.5cm
\put(-1.2,11){\rotate[r]{\epsffile{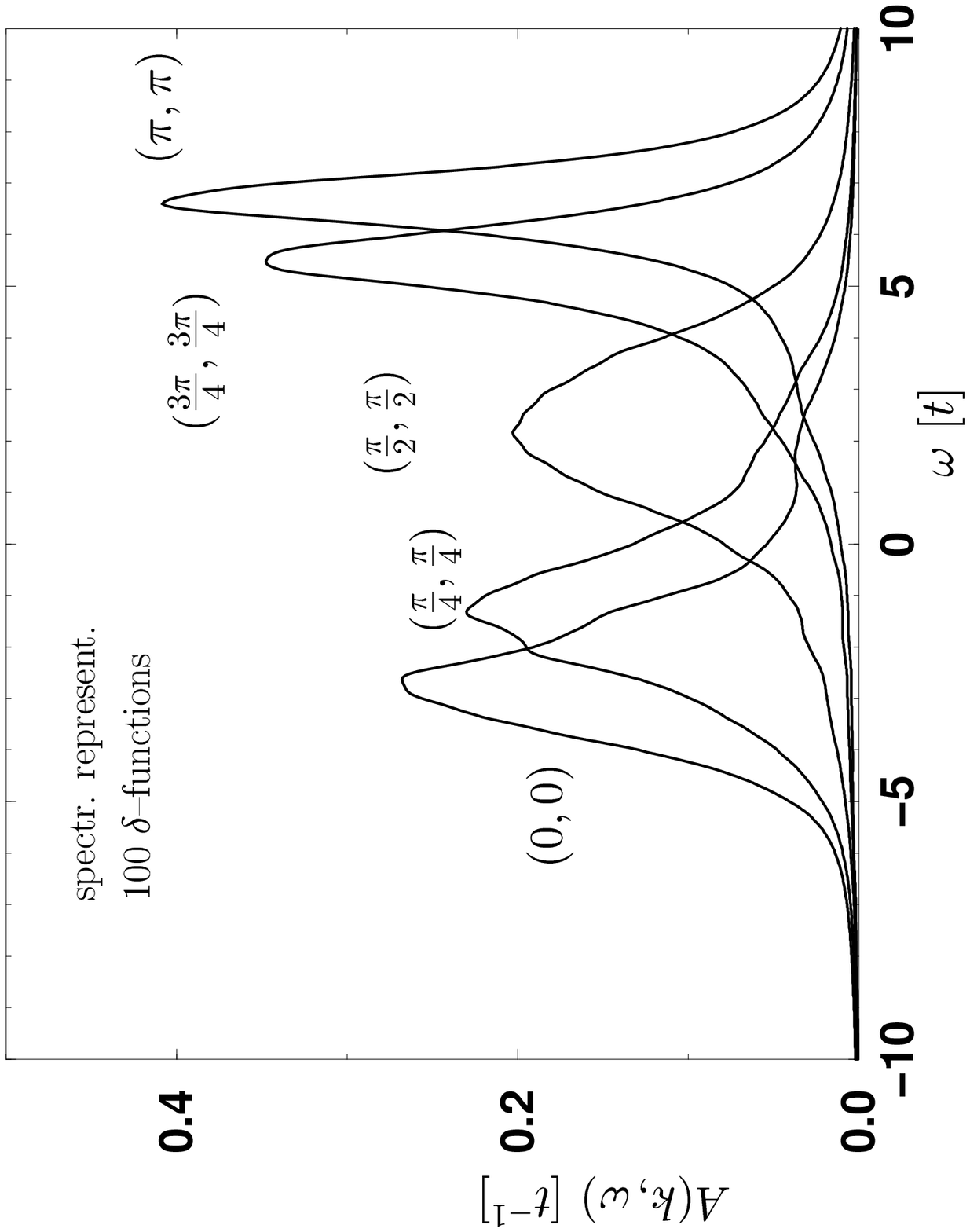}}}
\epsfxsize=7.5cm
\put(-1.2,5.5){\rotate[r]{\epsffile{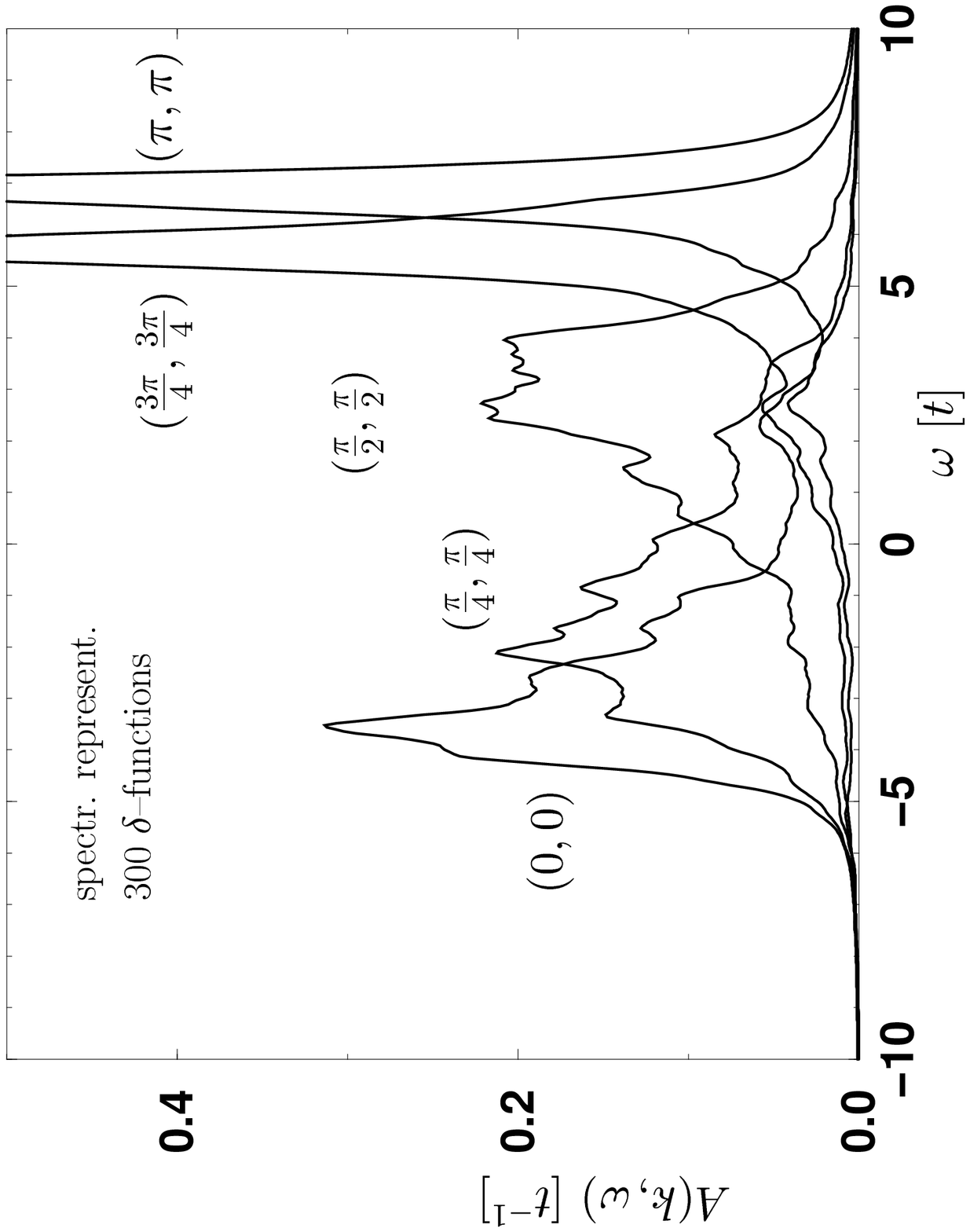}}}
\epsfxsize=7.5cm
\put(-1.2,0){\rotate[r]{\epsffile{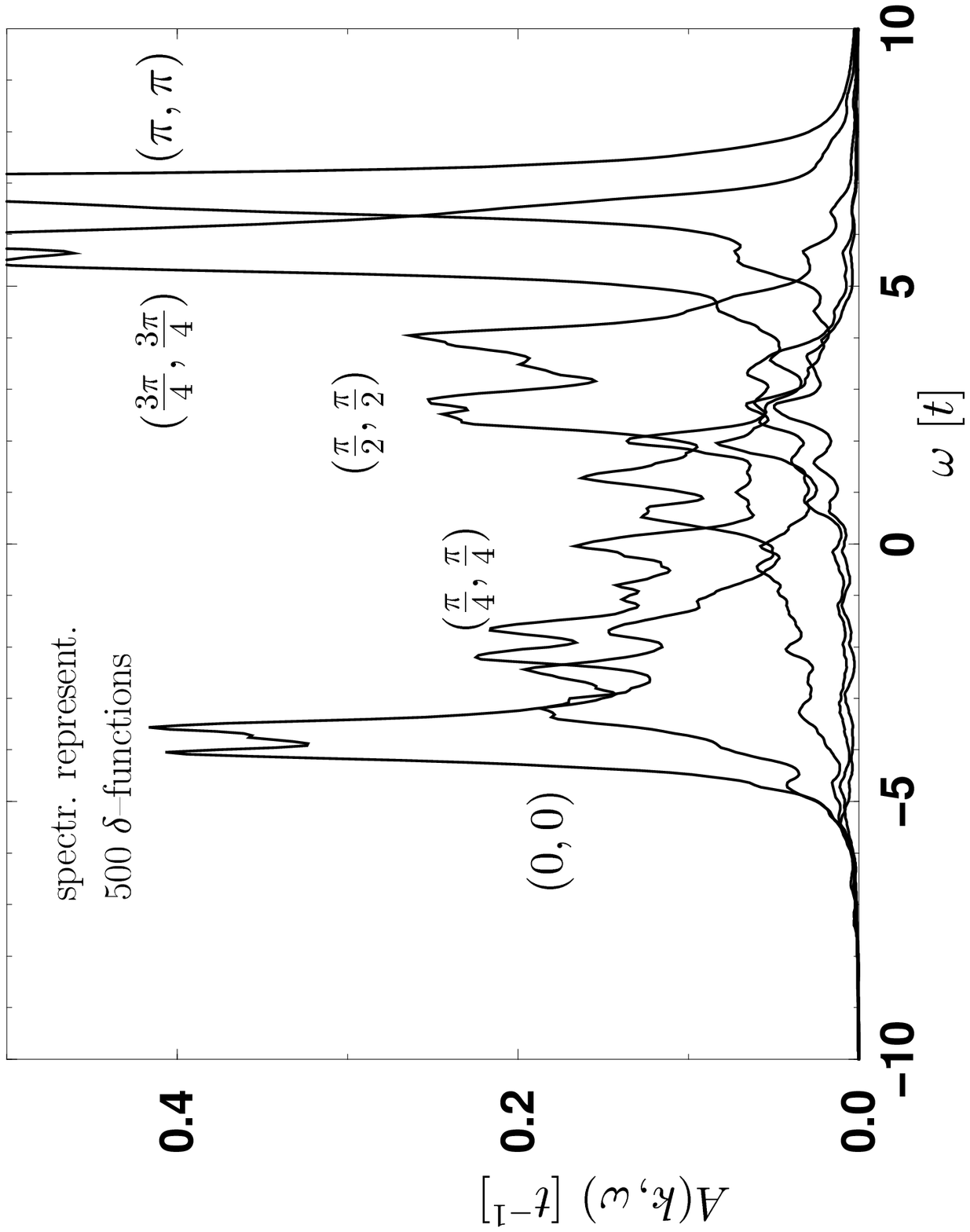}}}
\end{picture}
\caption{For the same parameters as in Fig. \protect\ref{fig:1} the ${\bf k}$
dependent spectral function along the $(1,1)$ direction is
plotted. (a) shows the result from the Matsubara technique whereas
(b-d) contain the result for $N_{max} = 100,300,500$
respectively.  
}
\label{fig:2}
\end{figure}

In Figs. \ref{fig:1} and \ref{fig:2} we compare the density of states
for a non-self-consistent calculation ( Eqs. (\ref{nsc1})--(\ref{nsc4}))
on an 8x8 lattice. The temperature was chosen to $0.55[t]$ and the
attractive interaction $U=-4[t]$ which is half the bandwidth of the 
non-interacting system. The chemical potential was fixed to $\mu=-2[t]$
which led to an electron density of $n\approx 0.7$ at this 
temperature. 
For the spectral
representation technique we used a frequency range of (see
Eq. (\ref{eq:grid2})) $\omega_{min} = -24[t] < \omega < \omega_{max} =
24[t]$, the parameter $\alpha$ in Eq. (\ref{eq:grid2}) was chosen
to be $\alpha = 2$ and we discuss the effect of a different number of
$\delta$-functions $N_{max}$. In Fig. \ref{fig:1}a we compare the
density of states as obtained with Matsubara technique and 
a numerically exact analytic continuation onto the real axis
\cite{marsiglio88} which is possible only for the non-self-consistent
calculation 
with a calculation for $N_{max}=300$. At $\omega - \mu =
2[t]$ there is a remnant of the logarithmic singularity which occurs
in the middle of the band of a non-interacting 2D system. Below that,
around $\omega=0$ and below the chemical potential clear correlation
effects can be seen which lead to additional states at $\omega <
0$. In Fig. \ref{fig:1}b we compare for the same parameters
different numbers $N_{max}$ of $\delta$-peaks. For $N_{max}=100$ the
correlation effects around the chemical potential are not clearly
visible whereas for   $N_{max}=300$ they are clearly
present. Increasing $ N_{max}$ up to $500$ does not alter the
picture.

In Fig. \ref{fig:2} we calculate ${\bf k}$ dependent quantities
along the $(1,1)$ direction. The results from the Matsubara technique 
show that there is a strong incoherent broadening
of the former quasiparticle peak around $k_F$ and for $k < k_F$ due to
correlations.
Along the diagonal the Fermi wave-vector is bracketed by
 $(\pi/4,\pi/4) < k_F <
(\pi/2,\pi/2)$  Already the
calculation with $N_{max}=100$ in Fig. \ref{fig:2}b resolves the
incoherent broadening but does fail to give further details which are
clearly visible in the calculation for Fig. \ref{fig:2}c for
$N_{max}=300$  and in Fig. \ref{fig:2}d for
$N_{max}=500$.

\begin{figure}
\unitlength1cm
\epsfxsize=10cm
\begin{picture}(7,16)
\put(-2.2,7){\rotate[r]{\epsffile{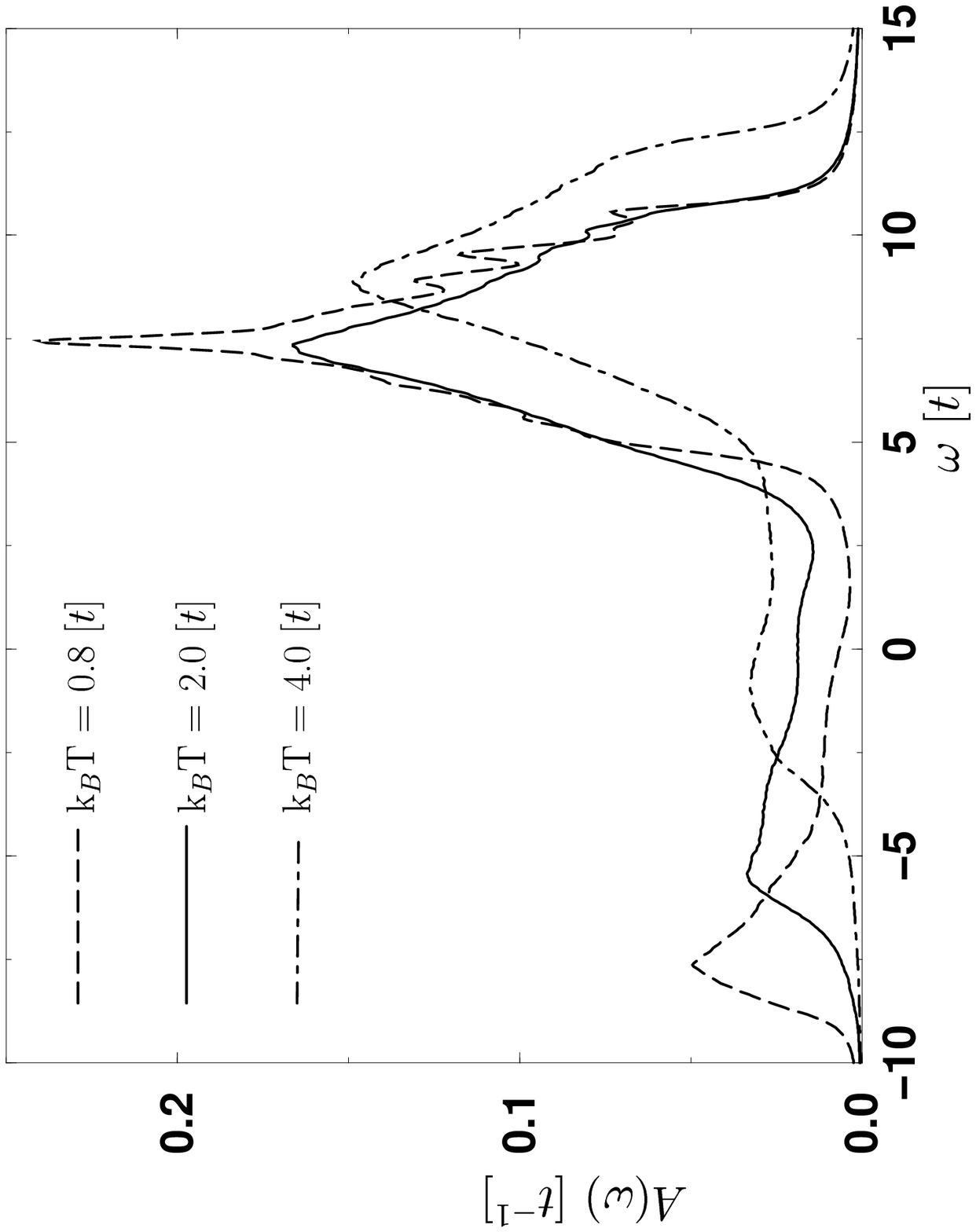}}}
\epsfxsize=10cm
\put(-2.2,0){\rotate[r]{\epsffile{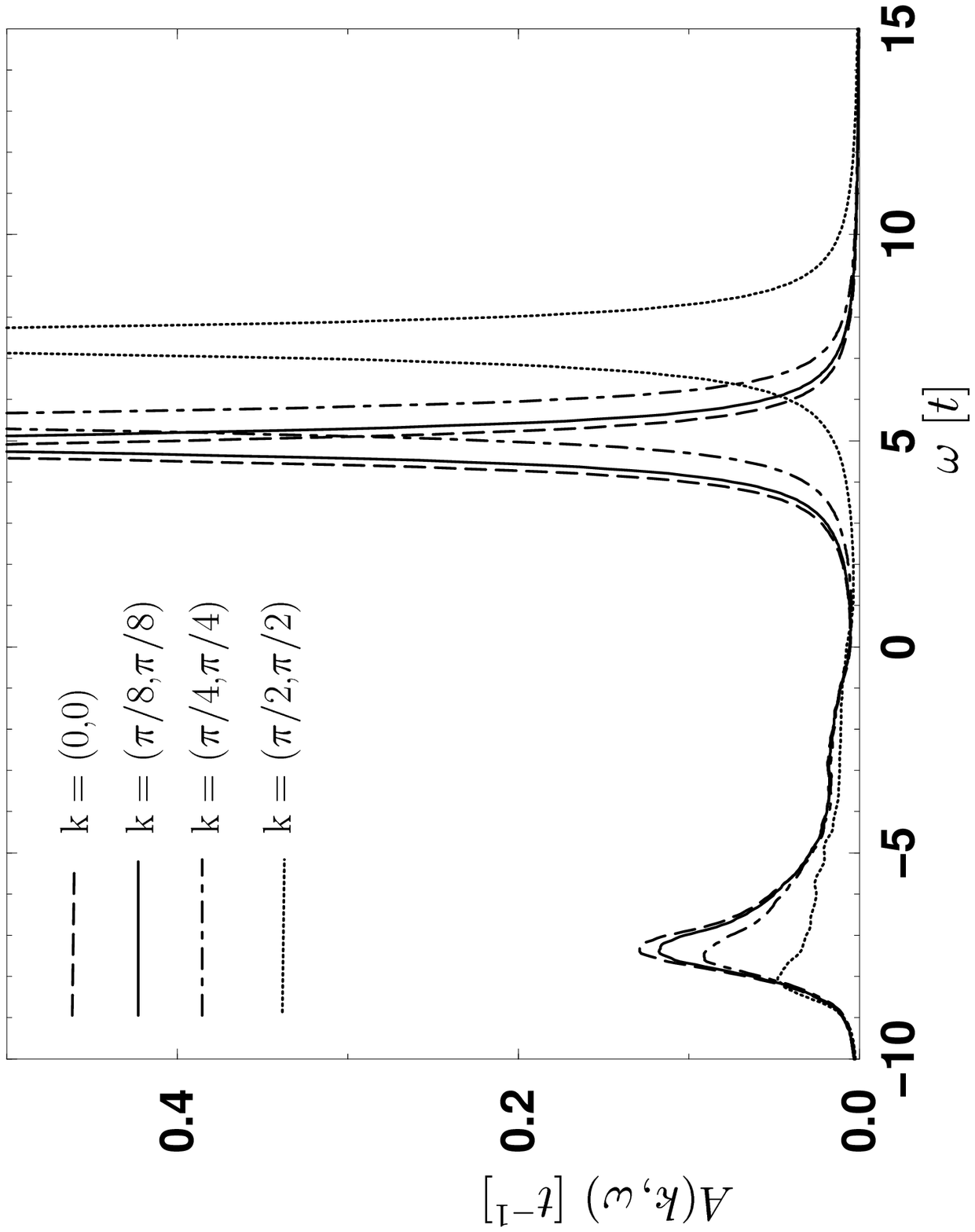}}}
\end{picture}
\caption{For a 16x16 cubic 2D lattice, $U=-8[t]$ and $n = 0.2$
the density of states is plotted for three different temperatures. the
chemical potential $\mu$ has been adjusted as a function of
temperature to keep $n$ constant. 
$\omega=0$ is the position of the chemical potential.
(a) shows the
density of states for the three temperatures $k_BT = 4.0[t]$
(dot--dashed line), $k_BT = 2.0[t]$
(full line) and $k_BT = 0.8[t]$ (dashed line). (b) shows for the
lowest temperature the ${\bf k}$-dependent spectral function along the
$(1,1)$ direction. The results were obtained with a non-self-consistent
calculation.
}
\label{fig:3}
\end{figure}

\begin{figure}
\unitlength1cm
\epsfxsize=10cm
\begin{picture}(7,8.5)
\put(-2.2,0){\rotate[r]{\epsffile{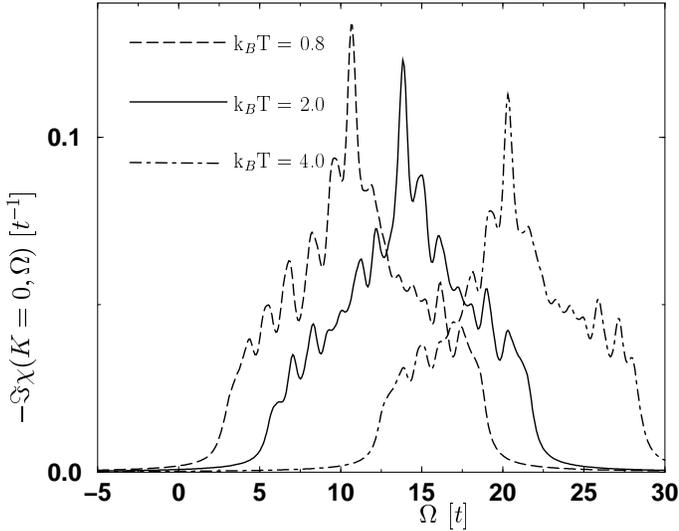}}}
\end{picture}
\caption{For the same parameters as in Fig. \protect\ref{fig:3} the
imaginary part of the
susceptibility for pairs of electrons with total momentum $K=0$ is
shown as obtained from a non-self-consistent calculation.
}
\label{fig:4}
\end{figure}

\begin{figure}
\unitlength1cm
\epsfxsize=10cm
\begin{picture}(7,8.5)
\put(-2.2,0){\rotate[r]{\epsffile{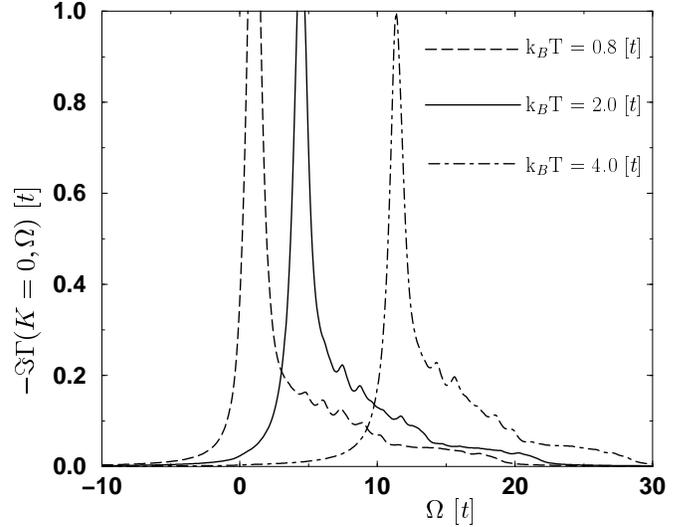}}}
\end{picture}
\caption{For the same parameters as in Fig. \protect\ref{fig:3} the
imaginary part of the
vertex function $\Gamma({\bf K=0},\Omega)$ for pairs of electrons with
total momentum $K=0$ is shown as it results from a non-self-consistent
calculation.  
}
\label{fig:5}
\end{figure}

\begin{figure}
\unitlength1cm
\epsfxsize=10cm
\begin{picture}(7,8.5)
\put(-2.2,0){\rotate[r]{\epsffile{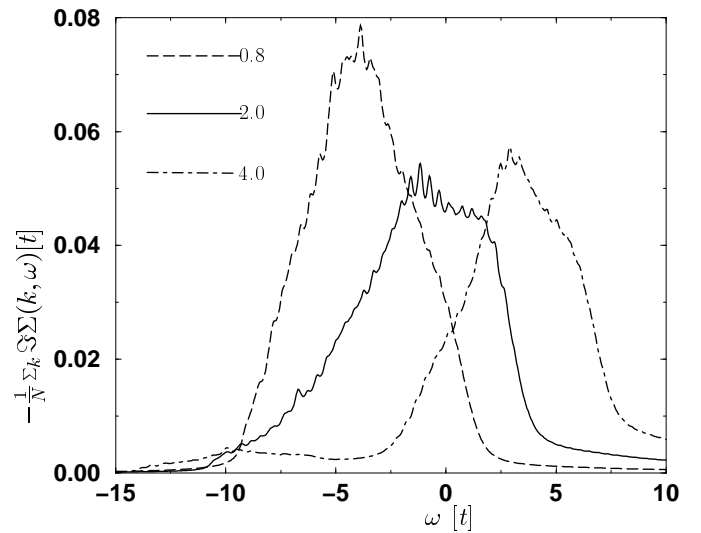}}}
\end{picture}
\caption{For the same parameters as in Fig. \protect\ref{fig:3} the
imaginary part of the
${\bf k}$-averaged self-energy is shown as it results from a 
non-self-consistent calculation.  
}
\label{fig:6}
\end{figure}

\begin{figure}
\unitlength1cm
\epsfxsize=10cm
\begin{picture}(7,16)
\put(-2.2,7){\rotate[r]{\epsffile{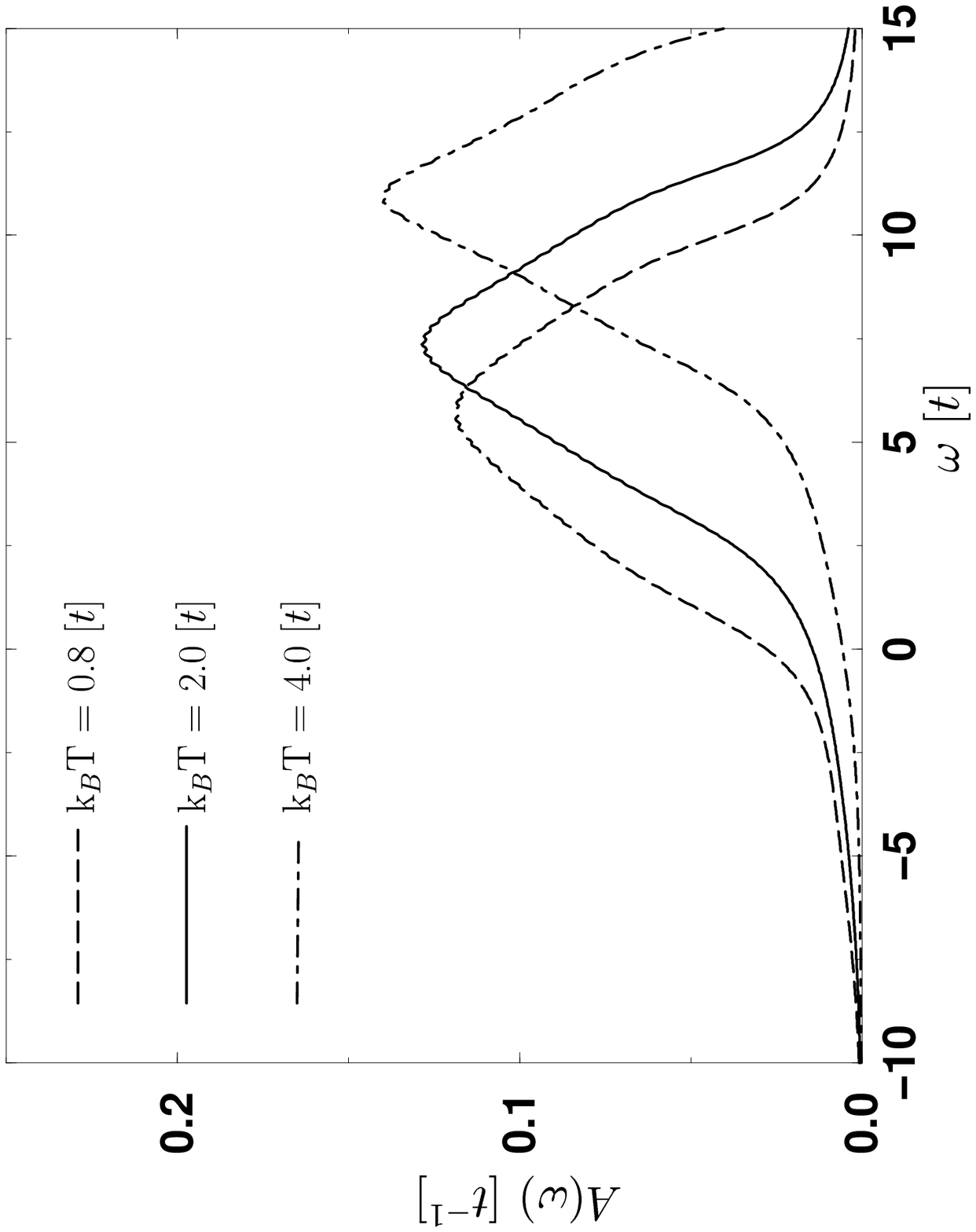}}}
\epsfxsize=10cm
\put(-2.2,0){\rotate[r]{\epsffile{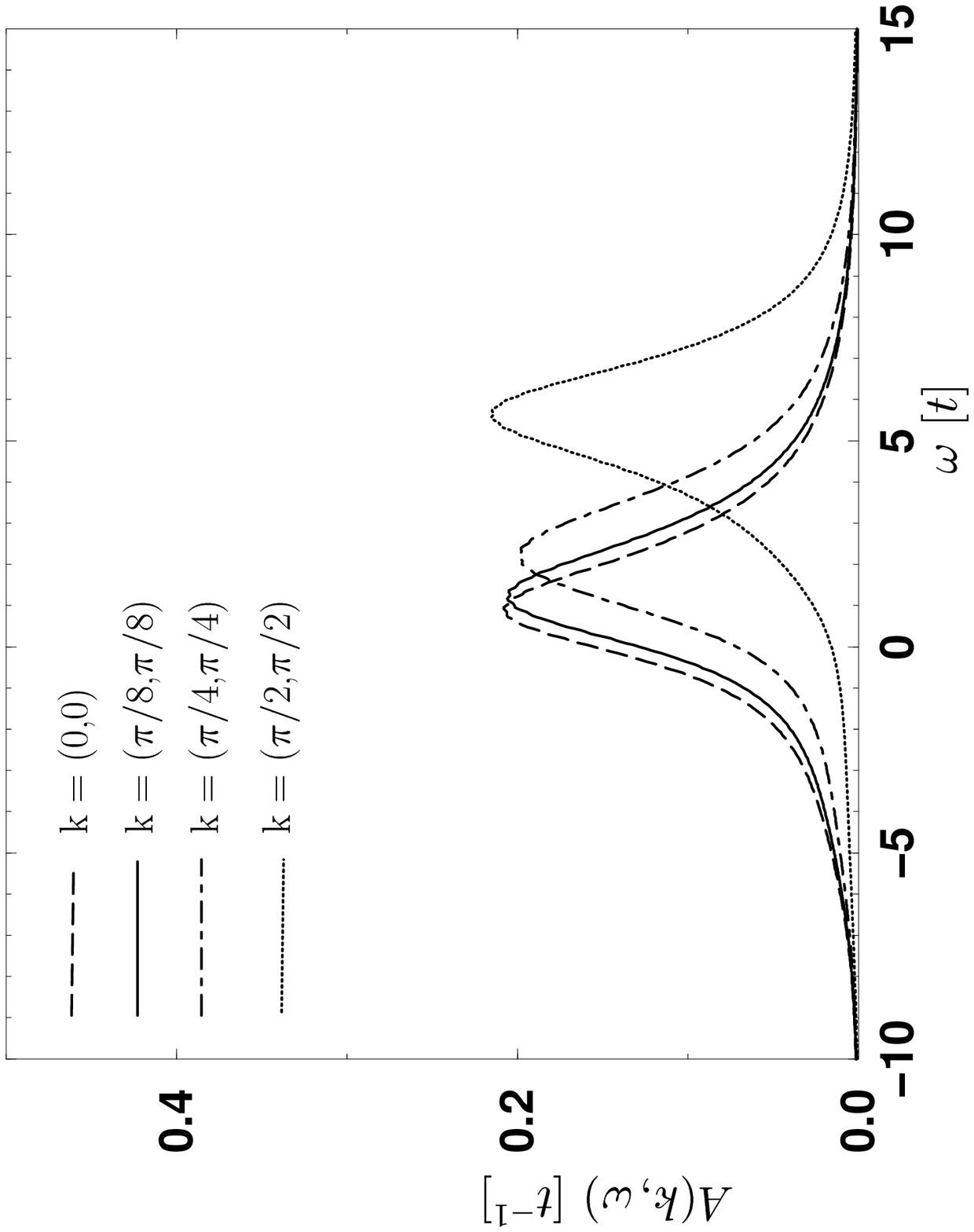}}}
\end{picture}
\caption{For the same parameters as in Fig. \protect\ref{fig:3} the density of
states resulting from a fully self-consistent calculation is
plotted in (a), while (b) shows the ${\bf k}$-dependent spectral
function. 
}
\label{fig:7}
\end{figure}

\begin{figure}
\unitlength1cm
\epsfxsize=10cm
\begin{picture}(7,8.5)
\put(-2.2,0){\rotate[r]{\epsffile{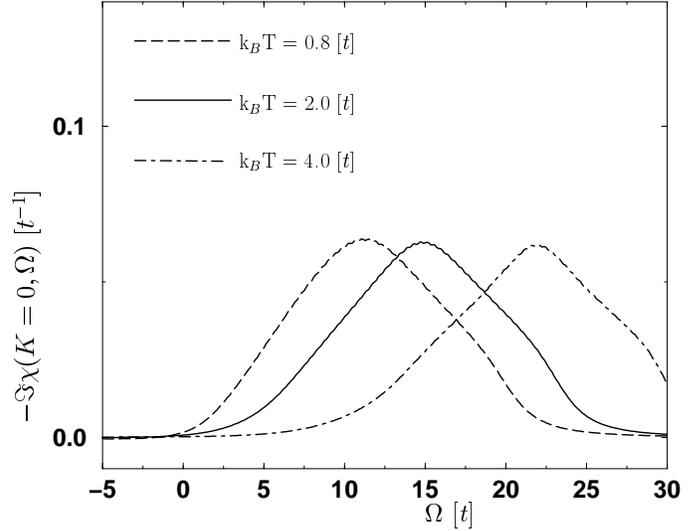}}}
\end{picture}
\caption{For the same parameters as in Fig. \protect\ref{fig:3} the
imaginary part of the
susceptibility for pairs of electrons with total momentum $K=0$ is
shown as it results from a fully self-consistent calculation.
}
\label{fig:8}
\end{figure}

\begin{figure}
\unitlength1cm
\epsfxsize=10cm
\begin{picture}(7,8.5)
\put(-2.2,0){\rotate[r]{\epsffile{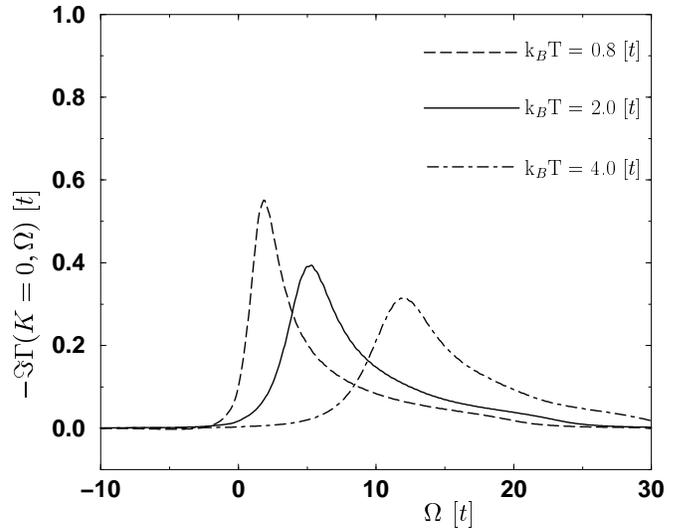}}}
\end{picture}
\caption{For the same parameters as in Fig. \protect\ref{fig:3} the
imaginary part of the
vertex function $\Gamma({\bf K=0},\Omega)$ for pairs of electrons with
total momentum $K=0$ is shown as it results from a fully
self-consistent calculation. 
}
\label{fig:9}
\end{figure}

\begin{figure}
\unitlength1cm
\epsfxsize=10cm
\begin{picture}(7,8.5)
\put(-2.2,0){\rotate[r]{\epsffile{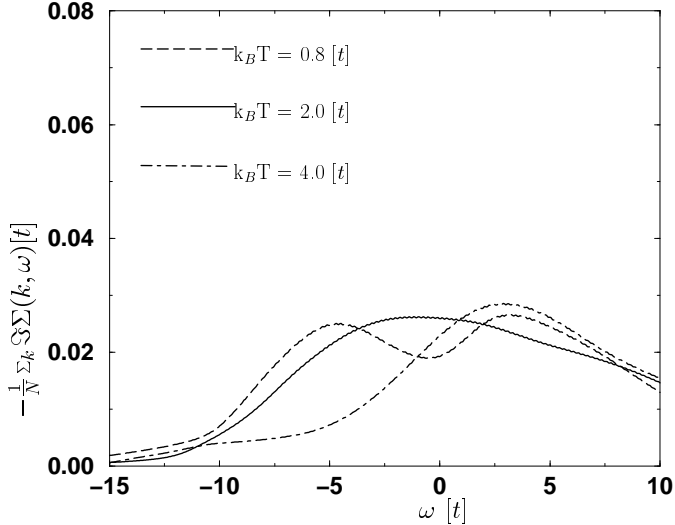}}}
\end{picture}
\caption{For the same parameters as in Fig. \protect\ref{fig:3} the
imaginary part of the
${\bf k}$-averaged self-energy is shown as it results from a
fully self-consistent calculation. 
}
\label{fig:10}
\end{figure}

In order to demonstrate the strength of our method we discuss in the
following (Figs. \ref{fig:3} - \ref{fig:10}) some aspects of the
temperature dependence of the correlation functions obtained for a
16x16 lattice with $N_{max}=300$, $\omega_{min} = -32[t]$ and
$\omega_{max} = 32[t]$ (see Eq. (\ref{eq:grid2})). The 
strength of the attractive
interaction for these figures is $U=-8[t]$, which is equal to the
bandwidth of the non-interacting system. The particle number was
chosen to be $n=0.2$ ($n=1$ would correspond to half filling) and the
chemical potential was adjusted as a function of temperature in order
to keep the particle number $n$ constant. In
Figs. \ref{fig:3}-\ref{fig:6} non-self-consistent results according to
Eqs. (\ref{nsc1})-(\ref{nsc4}) are presented whereas in
Figs. \ref{fig:7}-\ref{fig:10}, results from a fully self-consistent
calculation are presented.

Fig. \ref{fig:3}a shows the k-integrated density of states. With
decreasing temperature a gap occurs. The density at higher
temperatures results from the one--particle continuum and the density
for low energies results from pairs in a two--particle bound
state. Fig. \ref{fig:3}b shows that for $k_BT=0.8[t]$ also the
k-dependent spectral function consists of two parts especially around
$k=k_F$, $k_F \approx (\pi/8,\pi/8)$.

Fig. \ref{fig:4} shows the imaginary part of the susceptibility for a
total momentum of the pair of $K=0$ which is -- as it should be -- the
non-interacting density of states where the energy has been stretched by
a factor of 2. It is simply shifted according to the shift of the
chemical potential with temperature. Due to the low density in the
example we have chosen the chemical potential does not enter in the
one--particle continuum. This is the case for higher densities where
the imaginary part of the susceptibility changes sign at zero.

Fig. \ref{fig:5} shows the imaginary part of the vertex function
$\Gamma(K=0,\Omega)$. The strong peak corresponds to a true bound
state and with decreasing temperature, the chemical potential drifts
towards the bound state indicating Bose condensation of
non-interacting pairs into the bound state.

In Fig. \ref{fig:6} we show the k-averaged imaginary part of the 
self-energy $\Sigma(k,\omega)$. It mainly shifts with the chemical
potential but does not otherwise show a large temperature 
dependence.

In the following figures we show results from a self-consistent
calculation which have to be compared with the results from the
non-self-consistent calculation. Fig. \ref{fig:7}a shows the density
of states. Even though it looks similar for $k_BT=4[t]$ there is strong
difference at lower temperatures. The gap is no longer present due
to the self-consistent procedure. Also for the k-dependent spectral
functions (for  $k_BT=0.8[t]$ in Fig. \ref{fig:7}b) there is no
splitting into two parts.

Fig. \ref{fig:8} shows the imaginary part of the susceptibility. It is
no longer an image of the non-interacting density of states. 
Note that it becomes negative for $\Omega < 0$ at the
lowest temperature considered here ($k_BT=0.8[t]$) which is barely
visible from the plot and indicates that the chemical potential is now
in the one particle continuum. Also the imaginary part of the vertex
function $\Gamma({\bf K},\Omega)$, which is plotted for ${\bf K} = 0$
in Fig. \ref{fig:9}, changes sign at $\Omega = 0$, although this
seems to happen in the plot (Fig. \ref{fig:9}) for $\Omega < 0$,
which is just an artifact of the broadening of the $\delta$-functions
which had to be applied in order to plot a spectral quantity. The peak
in $\Im\Gamma({\bf K}=0,\Omega)$ no longer corresponds to a bound
state; it is a two particle resonance in the one-particle
continuum.

Also the self-energy whose imaginary part is (averaged over ${\bf k}$)
plotted in Fig. \ref{fig:10} is strongly altered due to
self-consistency. Compared to the non-self-consistent part in
Fig. \ref{fig:6} it is strongly decreased in magnitude and starts 
at low temperatures  to develop a minimum at $\omega = 0$
indicating the appearance of Fermi liquid like properties.

\section{Conclusion}
Using the example of the attractive Hubbard model we have evaluated the 
ladder diagrams of the T-matrix, and we have demonstrated
that our numerical method, which works entirely along the real frequency
axis, enables us to accurately calculate spectral properties. 
We should point out that we did not show results for the 
lowest temperatures we were
able to reach. In fact we can decrease the temperature for the
calculations of Figs. \ref{fig:3} - \ref{fig:10} by an additional two
orders of magnitude without reaching numerical instabilities. However
these results and especially their physical interpretation are not
the main subject of the current paper. Here we have described in detail the
numerical method and discussed its applicability to solve different
problems of correlated quantum systems.

\acknowledgements
We acknowledge financial support from the BMBF, Germany. M.L. further
thanks R.~J.~Gooding for pointing out the importance of the T-matrix
calculation and for stimulating discussions. 

%\bibliographystyle{unsrt}
%\bibliography{litf}

\begin{thebibliography}{10}

\bibitem{zubarev60}
D.~N. Zubarev.
\newblock {\em Usp.~Fiz.~Nauk (71)71 [Secs 25, 31]}, (1960).

\bibitem{tyablikov67}
S.~V. Tyablikov.
\newblock {\em Methods in the Quantum theory of Magnets}.
\newblock Plenum Press, New York, (1967).

\bibitem{matsubara55}
T.~Matsubara.
\newblock {\em Prog.~Theor.~Phys. {\bf 14}, 351}, (1955).

\bibitem{schrieffer63}
{J.~R.~Schrieffer, D.~J.~Scalapino and J.~W.~Wilkins}.
\newblock {\em Phys.~Rev.~Lett. {\bf 10} 336}, (1963).

\bibitem{scalapino66}
{D.~J.~Scalapino, J.~R.~Schrieffer, and J.~W.~Wilkins}.
\newblock {\em Phys.~Rev. {\bf 148} 263}, (1966).

\bibitem{owen71}
{C.~S.~Owen and D.~J.~Scalapino}.
\newblock {\em Physica (Amsterdam) {\bf 55} 691}, (1971).

\bibitem{bergmann73}
{G.~Bergmann and D.~Rainer}.
\newblock {\em Z. Physik {\bf 263} 59}, (1973).

\bibitem{rainer74}
{D.~Rainer and G.~Bergmann}.
\newblock {\em J. Low Temp. Phys. {\bf 14} 501}, (1974).

\bibitem{vidberg77}
J.~W.~Serene H.~J.~Vidberg.
\newblock {\em J. of Low Temp. Phys. {\bf 29}(3/4) 179}, (1977).

\bibitem{silver90}
{R.~N.~Silver, J.~E.~Gubernatis, D.~S.~Sivia, M.~Jarrell}.
\newblock {\em Phys.~Rev.~Lett. {\bf 65}, 496}, (1990).

\bibitem{randeria89}
M.~Randeria et~al.
\newblock {\em Phys. Rev. Lett. {\bf 62} 981}, (1989).

\bibitem{micnas95}
{R.~Micnas, M.H.~Pedersen, S.~Schafroth, T.~Schneider, J.J.~Rodriguez-Nunez,
  H.~Beck}.
\newblock {\em Phys.~Rev.~B {\bf 52} 16223}, (1995).

\bibitem{janko97}
{B.~Janko, J.~Maly, K.~Levin}.
\newblock {\em Phys.~Rev.~B {\bf 56} R11407}, (1997).

\bibitem{laplace}
A cleaner mathematical definition allows to define the frequency
dependent Green function as the complex Laplace transform of the
thermal expectation value of the commutator between two operators
$\langle [C(t-t'),B]_{\mp} \rangle$.

\bibitem{fetwal}
J.~D.~Walecka A.~L.~Fetter.
\newblock {\em Quantum Theory of Many-Particle Systems}.
\newblock McGraw-Hill, (1971).

\bibitem{AGD}
I.~E.~Dzyaloshinski A.~A.~Abrikosov, L.~P.~Gorkov.
\newblock {\em Methods of quantum field theory in statistical physics}.
\newblock Dover, (1975).

\bibitem{marsiglio88}
{F.~Marsiglio, M.~Schossmann and J.~P.~Carbotte}.
\newblock {\em Phys.~Rev.~B {\bf 37} 4965}, (1988).

\bibitem{fresard92}
{R.~Fresard, B.~Glaser, P.~W\"olfle}.
\newblock {\em J.~Phys.~Cond.~Mat. {\bf 4} 8565}, (1992).

\bibitem{kyung98}
{Bumsoo~Kyung, E.G.~Klepfish, P.E.~Kornilovitch}.
\newblock {\em Phys.~Rev.~Lett. {\bf 80}(14) 3109}, (1998).

\bibitem{letz98}
{M.~Letz, R.~J.~Gooding}.
\newblock {\em J. Phys. Cond. Mat. {\bf 10} (31) 6931-6951}, (1998).

\bibitem{haussmann93}
{R.~Haussmann}.
\newblock {\em Z.~Phys.~B {\bf 91} 291}, (1993).

\bibitem{thoulesskrit}
{D.~J.~Thouless}.
\newblock {\em Annals~of~Phys. {\bf 10} 553}, (1960).

\bibitem{svr}
{S.~Schmitt--Rink, C.~M.~Varma, A.~E.~Ruckenstein}.
\newblock {\em Phys.~Rev.~Lett. {\bf 63}, 445}, (1989).

\bibitem{serene89}
{J.~W.~Serene}.
\newblock {\em Phys.~Rev.~B {\bf 40}(16) 10873}, (1989).

\end{thebibliography}

\end{document}